\documentclass[apjl]{emulateapj}
\usepackage{apjfonts,mathptmx}

\def\xmm {\emph{XMM-Newton}}

\def\src {CXOU\,J010043.1--721134}

\begin{document}


\title{\emph{XMM-Newton} observations of \src: the first deep look at the soft X--ray emission of a magnetar\altaffilmark{1}}

\author{A.~Tiengo, P.~Esposito\altaffilmark{2}, and S.~Mereghetti}
\email{tiengo@iasf-milano.inaf.it}
\affil{INAF - Istituto di Astrofisica Spaziale e Fisica Cosmica Milano,\\
          Via Edoardo Bassini 15, 20133 Milano, Italy}

\altaffiltext{1}{Based on observations obtained with
\emph{XMM-Newton}, an ESA science mission with instruments and
contributions directly funded by ESA Member States and NASA.}
\altaffiltext{2}{Universit\`{a} di Pavia, Dipartimento di Fisica
Nucleare e Teorica and INFN-Pavia,
              via Agostino Bassi 6, 27100 Pavia, Italy}

\shortauthors{A. Tiengo et al.}
\shorttitle{\xmm\ observations of \src}

\journalinfo{The Astrophysical Journal Letters, in press}
\submitted{Received 2008 February 11; accepted 2008 May 14}

\begin{abstract}
We present the analysis of six \xmm\ observations of the Anomalous
X--ray Pulsar \src, the magnetar candidate characterized by the
lowest interstellar absorption.
In contrast with all the other magnetar candidates, its X--ray
spectrum cannot be fit by an absorbed power-law plus blackbody
model. The sum of two (absorbed) blackbody components with
$kT_1=0.30\pm0.02$ keV and $kT_2=0.7\pm0.1$ keV gives an
acceptable fit, and the radii of the corresponding blackbody
emission regions are $R^\infty_{\rm BB1}=12.1^{+2.1}_{-1.4}$ km
and $R^\infty_{\rm BB2}=1.7^{+0.6}_{-0.5}$ km. The former value is
consistent with emission from a large fraction of a neutron star
surface and, given the well known distance of \src, that is located
in the Small Magellanic Cloud, it provides the most constraining
lower limit to a magnetar radius ever obtained. A more physical
model, where resonant cyclotron scattering in the magnetar
magnetosphere is taken into account, has also
been successfully applied to this source.

\end{abstract}



%

\section{Introduction}

The Anomalous X--ray Pulsars \citep[AXPs; see][for a recent
review]{magnetarsSandro} were initially identified as a subclass of
accreting X--ray pulsars.
Their much softer X--ray
spectrum and the lack of a bright optical counterpart distinguished
them from the vast majority of X--ray pulsars, which are neutron
stars accreting in high mass X--ray binary systems. AXPs have
rotation periods of several seconds and show a secular spin-down on
timescales of 10$^3$--10$^5$ years, but their rotational energy loss
is smaller than their X--ray luminosity, excluding the possibility
that they are rotation-powered, like radio pulsars. It is generally
believed that the AXPs, as well as another small class of
high-energy sources with similar properties, the Soft Gamma-ray
Repeaters (SGRs), are magnetars, i.e. neutron stars powered by their
extremely high magnetic field
\citep[$\sim$10$^{15}$ G;][]{duncan92,thompson96}.

The soft X-ray (1--10 keV)  spectra of magnetars cannot be
adequately fit with single component models whenever data with good
count statistics are available. Successful fits are instead obtained
by a two component model consisting
of a steep power-law (photon index $\sim$3--4) and a blackbody
($kT\sim0.5$ keV).
Some attempts have been done, also based on
phase-resolved spectroscopy, to attribute the two components to
physically distinct processes \citep[e.g.][]{tmt05}, but no
particularly compelling interpretations could be obtained.
One problem of
this model is that it tends to give best-fit values of the
interstellar absorption higher than those independently estimated in
other ways \citep[e.g.][]{durant06}. Another problem is that the
power-law component cannot be extrapolated at lower energies without
exceeding the flux of the near infrared (NIR) and optical
counterparts \citep[e.g.][]{hulleman04}. Drastic, and possibly
unphysical, cut-offs in the power-law component are required to
match the low optical/NIR fluxes.

In some AXPs
good spectral fits are obtained with the sum of
two blackbody components with different temperatures. Since this
model does not suffer of the problems described above,
it is usually preferred to the power-law plus blackbody model
\citep[e.g.][]{halpern05}. However, also this model is only
phenomenological and it is inadequate to represent the non-thermal
phenomena that are expected to occur in the highly magnetized
magnetosphere of magnetars \citep[e.g.][]{lyutikov06}. More physical
models of the X--ray spectra, including the effects of the strong
magnetic field and charged currents,
have recently been developed and successfully applied
to a sample of magnetar candidates
\citep[][]{fernandez07,guver07,RCSnanda}.

From a purely observational point of view, it has not been possible
to discriminate between the different models reproducing the
magnetar X--ray spectra.
This is mainly due to the low sensitivity of hard X--ray detectors
above $\sim$10 keV and to the large uncertainties in the fits
introduced by the high interstellar absorption, that severely
suppresses the flux below $\sim$1 keV. Being young neutron stars
born from  massive progenitors, all the Galactic magnetars are
located in highly absorbed regions of the Galactic plane. All of
them have column densities N$_{\rm{H}}$ ranging from
$\sim$$5\times10^{21}$ to $\sim$$10^{23}$ cm$^{-2}$.
The two known magnetars in the Magellanic Clouds, being considerably
less absorbed, offer the possibility to better constrain the spectra
in the low energy   range. The study of SGR~0526--66, located in the
Large Magellanic Cloud, is complicated by the presence of the
surrounding supernova remnant N49, which is particularly bright in
soft X--rays \citep{kulkarni03}. Here we concentrate therefore on
the spectral properties of \src\
\citep{lamb02,lamb03errata,mcgarry05}, the only known AXP in the
Small Magellanic Cloud (SMC).

\newpage
\section{Observations and data analysis}

The field containing \src\ was observed six times\footnote{Only
the data of the two first observations have already been
published \citep{lamb02,naze04,majid04,mcgarry05,nakagawa08};  we reanalyzed
them using more recent calibration files, in order to consistently
compare the results with those of the new observations.}  with the
\xmm\ satellite (see Table \ref{log}). Here we report the analysis
of the data collected by the EPIC instrument, which is composed by
one PN \citep{struder01} and two MOS X--ray cameras \citep{turner01}.
\begin{deluxetable}{ccccc}
  \tablecolumns{1}
\tablewidth{0pt}
 \tablecaption{\label{log} Log of the \emph{XMM-Newton} Observations of \src. The Pulse Periods
and Corresponding 1$\sigma$ Errors Are Also Reported.}
\tablehead{
\colhead{Obs.} & \colhead{Date} & \colhead{PN exposure} & \colhead{MOS exposure} & \colhead{Period}\\
\colhead{} & \colhead{} & \colhead{(ks)} & \colhead{(ks)} & \colhead{(s)}
}
\startdata
A & 17 Oct 2000 & 14 & 20 & 8.019(1)\phantom{1}\\
B & 20 Nov 2001  & 22 & 27&8.0193(9)\\
\,\,C\tablenotemark{a} & 27 Mar 2005  & \nodata & 24 & 8.0215(9)\\
D & 27 Nov 2005  & 14 & 17 & 8.022(1)\phantom{1}\\
E & 29 Nov 2005  & 13 & 16 & 8.022(1)\phantom{1}\\
F & 11 Dic 2005  & 9 & 16 & 8.022(2)\smallskip
\enddata
\tablenotetext{a}{The PN data were not considered, since the PN
instrument was operated with the filter wheel in closed position.}
\end{deluxetable}

\src\ was not the main target of the observations,
but, being at an off-axis angle of $\sim$$6\arcmin$, it was always well
inside the field of view of the EPIC instrument ($\sim$$15\arcmin$
 radius).
All the observations were performed with the medium optical blocking
filter and in Full Frame mode (time resolution of 73 ms and 2.6 s
for the PN and MOS, respectively), except for the first PN
observation, done in Extended Full Frame mode (time resolution of
200 ms). All the data were processed using the \xmm\ Science
Analysis Software (SAS version 7.1.0) and the calibration files
released in August 2007. The standard pattern selection criteria
(patterns 0--4 for PN and 0--12 for MOS) were adopted.

Source spectra were extracted for each observation
from circular regions with 25$\arcsec$ radius.
The background spectra were extracted from a
94$\arcsec$$\times$72$\arcsec$ box centered at
RA $=01^{\rm{h}}\,00^{\rm{m}}\,56\fs8$,
Dec. $=-72\degr\,11\arcmin\,33\arcsec$ and rotated such that it
intercepts no CCD gaps in any PN image. Response matrices and
ancillary files for each spectrum were produced using the SAS
software.

The spectra
were fitted to a set of models
(power-law, blackbody, blackbody plus power-law, and two
blackbodies, all modified by photoelectric absorption)
using XSPEC version 11.3.1. The single component models gave only
marginally acceptable fits in most observations, while better
results were obtained with the two component models. The best-fit
parameters of the latter models are reported in Table \ref{fits}.

\begin{deluxetable*}{cccccccccc}
  \tablecolumns{10}
\tablewidth{0pc}
 \tablecaption{Summary of the EPIC Spectral Results in the 0.1--10 keV Energy Range. Errors Are Given at the 90\% Confidence Level.\label{fits}} \tablehead{ \colhead{Observation} &
\colhead{Model\tablenotemark{a}} & \colhead{$N_{\rm
H}$\tablenotemark{b}} & \colhead{$\Gamma$}
& \colhead{PL norm.\tablenotemark{c}} & \colhead{$k_B T_{BB1}$} & \colhead{$R_{BB1}$\tablenotemark{d}} & \colhead{$k_B T_{BB2}$} & \colhead{$R_{BB2}$\tablenotemark{d}} & \colhead{$\chi^{2}_{r}$ (d.o.f.)} \\
\colhead{} & \colhead{} & \colhead{($10^{20}\rm cm^{-2}$)} & \colhead{} & \colhead{} & \colhead{(keV)} & \colhead{(km)} & \colhead{(keV)} & \colhead{(km)} & \colhead{}
}
\startdata
A
 & PL+BB  & $4.8^{+8.4}_{-2.1}$ & $1.3^{+1.1}_{-1.5}$ & $0.8^{+3.9}_{-0.7}$ & $0.35^{+0.02}_{-0.03}$ & $9.1^{+1.4}_{-0.9}$ & \nodata & \nodata & 1.18 (83)\phantom{1} \\
 & BB1+BB2  & $8.1^{+7.9}_{-4.0}$ & \nodata & \nodata & 0.27$^{+0.07}_{-0.09}$ & $12.9^{+15.7}_{-3.4}$ & 0.5$^{+0.3}_{-0.1}$& $3.6^{+3.5}_{-2.9}$ & 1.11 (83)\phantom{1} \\
B
 & PL+BB  & $4.3^{+3.0}_{-2.8}$ & $1.7^{+0.7}_{-0.5}$ & $2.0^{+5.7}_{-1.3}$ & $ 0.36^{+0.02}_{-0.03}$ & $8.5^{+1.3}_{-0.6}$ & \nodata & \nodata & 0.93 (115) \\
 & BB1+BB2  & $4.2^{+2.7}_{-2.1}$ & \nodata & \nodata & 0.31$\pm$0.03 & $11.0^{+2.6}_{-1.6}$ & 0.7$\pm$0.1& $1.6^{+0.9}_{-0.6}$ & 0.72 (115) \\
\,\,C\tablenotemark{e}
 & PL+BB  & $7.8^{+22}_{-7.0}$ & $\sim$2 & $<$16 & 0.34$^{+0.04}_{-0.05}$ & $9.6^{+2.8}_{-2.0}$ & \nodata & \nodata & 1.16 (107) \\
 & BB1+BB2  & $5.3^{+21.3}_{-4.4}$ & \nodata & \nodata & $<$0.39 & $<$11.6 & $\sim$1& $<$3.7 & 1.16 (107) \\
D
 & PL+BB  & $3.6^{+6.6}_{-3.5}$ & $1.6^{+0.7}_{-0.9}$ & $2.0^{+5.0}_{-1.6}$ & $ 0.36^{+0.02}_{-0.03}$ & 8.2$^{+1.0}_{-0.7}$ & \nodata & \nodata & 1.07 (83)\phantom{1} \\
 & BB1+BB2  & $<3.6$ & \nodata & \nodata & 0.35$^{+0.02}_{-0.03}$ & $8.9^{+0.9}_{-1.0}$ & 0.9$^{+0.4}_{-0.2}$& $0.7^{+0.7}_{-0.4}$ & 1.02 (83)\phantom{1} \\
E
 & PL+BB  & $4.2^{+7.2}_{-3.2}$ & $1.4^{+0.5}_{-1.2}$ & $1.5^{+5.0}_{-1.3}$ & $ 0.37^{+0.02}_{-0.03}$ & $8.3^{+1.1}_{-1.0}$ & \nodata & \nodata & 0.82 (84)\phantom{1} \\
 & BB1+BB2  & $4.4^{+3.7}_{-2.5}$ & \nodata & \nodata & 0.33$^{+0.04}_{-0.06}$ & $10.1^{+3.5}_{-1.7}$ & 0.8$^{+0.2}_{-0.3}$& $1.2^{+1.6}_{-0.7}$ & 0.75 (84)\phantom{1} \\
F
 & PL+BB  & $7.6^{+32.6}_{-5.6}$ & $1.6^{+2.0}_{-1.0}$ & 2.6$\pm$2.1 & $ 0.33^{+0.03}_{-0.06}$ & $9.9^{+2.6}_{-1.5}$ & \nodata & \nodata & 1.39 (65)\phantom{1} \\
 & BB1+BB2  & $6.6^{+6.4}_{-3.4}$ & \nodata & \nodata & 0.31$^{+0.04}_{-0.08}$ & $11.7^{+5.4}_{-2.3}$ & 0.8$^{+0.4}_{-0.3}$& $1.0^{+2.3}_{-0.6}$ & 1.30 (65)\phantom{1} \\
B+D+E+F\tablenotemark{f}
 & PL+BB  & $9.1^{+7.9}_{-3.8}$ & $2.0^{+0.5}_{-0.4}$ & $3.7^{+5.5}_{-1.9}$ & $0.34\pm0.02$ & $9.3^{+0.9}_{-0.7}$ & \nodata & \nodata & 1.75 (100) \\
 & BB1+BB2  & $6.3^{+2.0}_{-1.6}$ & \nodata & \nodata & $0.30\pm0.02$ & $12.1^{+2.1}_{-1.4}$ & $0.68^{+0.09}_{-0.07}$ & $1.7^{+0.6}_{-0.5}$ & \phantom{1}\phantom{1}1.14 (100)\smallskip
\enddata
\tablenotetext{a}{A free normalization factor has been introduced to
account for inaccurate flux reconstruction in the MOS spectra,
where, in most cases, the source is located on a CCD gap.}
\tablenotetext{b}{Assuming photoelectric absorption cross section
from \citet{balucinska92} and abundances from \citet{anders89}.}
\tablenotetext{c}{In units of 10$^{-5}$ ph. cm$^{-2}$ s$^{-1}$
keV$^{-1}$, at 1 keV.} \tablenotetext{d}{Assuming a distance of 60
kpc.} \tablenotetext{e}{Only MOS data.} \tablenotetext{f}{Only PN
data.}
\end{deluxetable*}

In order to check for flux variability, we have also
simultaneously fit the five available PN spectra and the two MOS
spectra for observation C with a double blackbody model with all
parameters linked to the same value and a variable normalization
factor.
From this analysis we can exclude significant ($>$3 $\sigma$) flux
variations larger than $\sim$20\% among the different \xmm\
observations.

\begin{figure}
\resizebox{\hsize}{!}{\includegraphics[angle=-90]{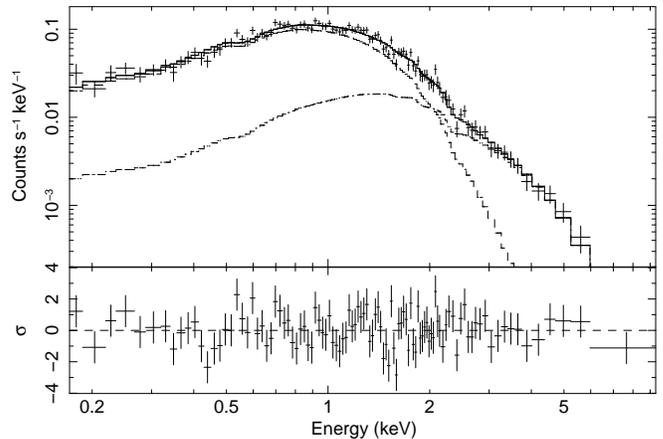}}
\caption{\label{spec} Cumulative PN spectrum of \src\ during
observations B, D, E, and F. The best-fit double blackbody model is
also shown together with the residuals and the separate contribution
of the two blackbody components (dotted and dash-dotted lines).}
\end{figure}

Since no significant differences in the spectral parameters are
detected and the calibration of the PN instrument has proven to be
very stable throughout the \xmm\ mission \citep[see,
e.g.,][]{kirsch05stab},
a cumulative spectrum of the PN data of observations B, D, E, and
F\footnote{Observation A has been excluded because it was taken in a
different operating mode, while
no PN data were available for observation C.} has also been
extracted.
The resulting net exposure time is 58 ks.
Only the double blackbody model gives an acceptable fit to the
cumulative spectrum with $kT_1\sim0.3$ keV and $kT_2\sim0.7$ keV
(see Fig.~\ref{spec} and Table \ref{fits}). The hydrogen column
density is in good agreement with the average value of N$_{\rm
H}=5.9\times10^{20}$ cm$^{-2}$ expected towards this region of
the SMC \citep{dickey90}. The observed flux in the
2--10 keV energy range
is $1.4\times10^{-13}$ erg cm$^{-2}$ s$^{-1}$,
corresponding to an unabsorbed
luminosity of $6.1\times10^{34}$ erg s$^{-1}$ for a distance of 60
kpc.
The double blackbody model gave
also the lowest $\chi^2$ values for the spectra of the single
observations, but in those cases it was not the only one compatible
with the data. In particular, the power-law plus blackbody model,
that is rejected with high confidence by the fit to the cumulative
spectrum, gave acceptable fits to all the single spectra.

In addition to the phenomenological models described above, we have
also fitted the cumulative spectrum with the magnetar spectral model
described in \citet{RCSnanda}. This model, originally proposed by
\citet{lyutikov06}, is based on cyclotron resonant scattering of
blackbody radiation in a twisted magnetosphere \citep{tlk02}.
Although the resulting $\chi^{2}_{r}$ (1.20 for 100 degrees of
freedom) is slightly worse than for the double blackbody model, the
fit is acceptable. The photoelectric absorption (N$_{\rm
H}=5\pm1\times10^{20}$ cm$^{-2}$) and blackbody temperature
($kT=0.32\pm0.08$ keV) are consistent with the values derived from
the double blackbody fit (N$_{\rm
H}=6.3^{+2.0}_{-1.6}\times10^{20}$ cm$^{-2}$ and
$kT=0.30\pm0.02$ keV for the cooler blackbody). The best-fit values
of the other spectral parameters
are a resonant scattering optical depth of
$\tau_{res}=1.2\pm0.2$
and a particle velocity of $\beta_T=0.48\pm0.12$. These parameters
are in the same range as the ones observed in the other magnetar
candidates \citep{RCSnanda}. Although
a direct information on the size of the emitting region cannot be
derived in the current version of this model, an approximate
estimate gives a radius similar to the one of the cooler component
in the double blackbody model.

\begin{figure}
\centering
\resizebox{\hsize}{!}{\includegraphics{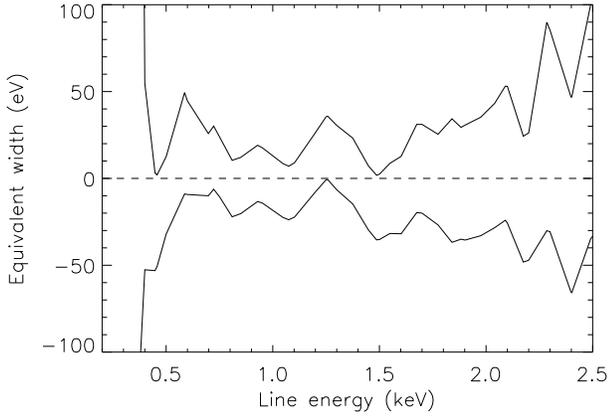}}
\caption{\label{lines} Upper limits (at 3 $\sigma$) to the
equivalent width of narrow lines (either in emission or absorption)
in the PN cumulative spectrum of \src\ during observations B, D, E,
and F.}
\end{figure}

By inspecting the residuals from the best-fit models, we found no
significant absorption or emission narrow-line features. We computed
upper limits on narrow lines' equivalent widths as a function of the
assumed line energy and width $\sigma_E$. This was done by adding
Gaussian components to the double blackbody model and computing the
allowed range in their normalization. The results for the
high-statistics cumulative PN spectrum are summarized in
Fig.\,\ref{lines}, where the plotted curves represent the
3\,$\sigma$ upper limits for $\sigma_E=0$ eV.

We performed a timing analysis to measure the source pulse period in
each data set. After correcting the   photon arrival times to the
Solar system barycenter,
we derived the best period values based on a  $Z_2^2$ periodogram
analysis \citep{buccheri83}. The resulting values are indicated in
Table \ref{log}.
Considering also the periods measured by \emph{Chandra}
\citep{mcgarry05}, a linear fit to the ten
values yields a period derivative
$\dot{P}=(1.9\pm0.1)\times10^{-11}$ s s$^{-1}$ ($\chi^2_{r}$ of 1.32
for 8 degrees of freedom).

Since observation D and E were performed only two days apart, we
tried to better constrain the spin-down rate through a
phase-coherent timing analysis of the two datasets. However, the
periods uncertainties during each observation
are too large to allow the prediction of the phase of the next
observation to better than a pulse cycle.

\begin{figure}
\centering
\resizebox{\hsize}{!}{\includegraphics{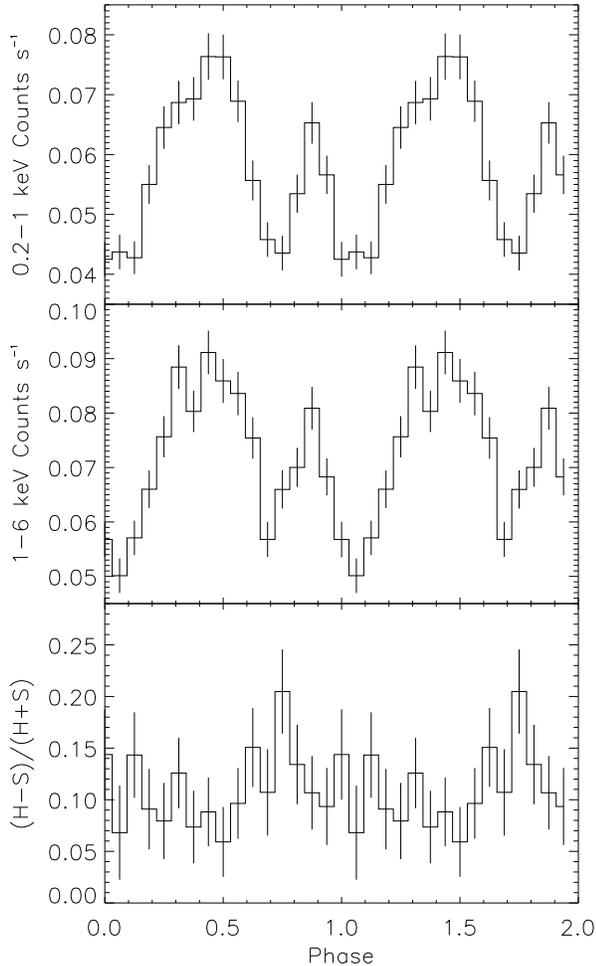}}
\caption{\label{profile} Background subtracted pulse profile of
\src\ in different energy ranges (as indicated in the panels) and
corresponding hardness ratio (PN data of observations B, D, E, F).}
\end{figure}

Many AXPs and SGRs are known to exhibit significant changes in their
pulse profiles \citep[e.g.][]{kaspi07,gogus02}. To search for
possible pulse shape variations in \src\ as a function of time, we
compared the folded light curves using a Kolmogorov-Smirnov test.
Taking into account the unknown relative phase alignment, all the
light curves are compatible with the same profile. We therefore
summed them after appropriate phase shifts\footnote{We selected the
shifts that maximized the Kolmogorov-Smirnov statistics comparing
subsequent observations.}.
The resulting pulse profiles in the soft (0.2--1 keV, $S$) and hard
(1--6 keV, $H$) energy ranges, together with their hardness ratio
(computed as $(H-S)/(H+S)$), are shown in Fig. \ref{profile}. This
analysis does not show any significant profile changes with energy.
The pulsed fraction\footnote{The pulsed fraction is defined as
$(C_{\rm{max}}-C_{\rm{min}})/(C_{\rm{max}}+C_{\rm{min}})$, where
$C_{\rm{max}}$ and $C_{\rm{min}}$ are the background-subtracted
count rates at the peak and at the minimum.} in the 0.2--6 keV energy
range is $(32\pm3)$\%.

\section{Discussion}
The new \xmm\ observations reported here indicate that \src\
continued to spin-dow at a rate of $\sim$$1.9\times10^{-11}$ s
s$^{-1}$, consistent with previous results \citep{mcgarry05}.
Although the errors on P are quite large, there is no evidence for
major torque changes. Also the pulse profile, spectral shape and
flux are consistent with no major changes, confirming that this AXP
is one of the most stable members of its class. This characteristic
allows us to sum up all the \xmm\ observations taken with the same
instrumental settings.

The cumulative PN spectrum of \src\ cannot be adequately fit by a
power-law plus blackbody model. This is the first magnetar for which
such a conclusion can be derived based only on the X--ray data,
thanks to the very low interstellar absorption.

A good fit is instead obtained with a double blackbody model. The
known distance of the SMC \citep[60 kpc,][]{keller06} allows a
precise measure
of the size of the emitting
region of the two blackbodies.
The $\sim$2 km radius of the region associated to the hotter
component is compatible with a hot spot on the neutron star surface.
The radius of the emitting region of the cooler blackbody,
12.1$^{+2.1}_{-1.4}$ km,
is consistent with a large fraction of the magnetar surface.
However, the strong pulsation below 1 keV (see Fig.~\ref{profile}),
where this component dominates
(see Fig.~\ref{spec}), indicates
that it cannot come from the whole neutron star surface. A similarly
large blackbody radius ($\sim$10 km) was also derived from the
spectrum of the AXP XTE~J1810--197 observed by \emph{ROSAT} before
the onset of its outburst in 2003 \citep{halpern05} and from
phase-resolved spectroscopy of \xmm\ observations of the same object
\citep{israel08}
 and of the AXP 1E~1048.1--5937 \citep{tmt05}. However,
in these cases, the less accurately known distance and the high
interstellar absorption produce large uncertainties on the emitting
region size.

Assuming that the thermal photons are produced on the neutron star
surface (and not, for instance, in the magnetosphere) and
considering that the blackbody is the most efficient thermal
emission process at a given temperature, the radius of the region
emitting the colder blackbody in \src\ is a firm lower limit to the
radius of the compact object. This limit is not large enough to
exclude any of the most popular equations of state for neutron
stars, but it is the most constraining lower limit ever obtained for
a magnetar.

Magnetar spectra are expected to be more complex than a double
blackbody. In fact most magnetar spectra cannot be fitted by such a
simple model, that underestimates their emission in the 5--10 keV
energy range. In all these cases a power-law tail in hard X--rays
($>$20 keV) has been detected \citep[see,
e.g.,][]{kuiper06,gotz06,leyder08} and it is likely responsible also
for the hard excess below 10 keV \citep{nakagawa08}. From a
theoretical point of view, the emission expected from a magnetar has
two components, with a thermal part directly from the surface, and a
non-thermal one due to emission reprocessed in the magnetosphere
\citep{lyutikov06,fernandez07}. We found that also \src\ can be fit
by a model of this kind \citep{RCSnanda}.

No compelling detections of spectral features in the persistent
X--ray emission of magnetars have been reported so far. \xmm\ and
\emph{Chandra} observations yielded strong upper limits (equivalent
width $\lesssim$10 eV) for 4U\,0142+61 \citep{juett02},
1E\,1048.1$-$5937 \citep{tmt05}, and 1E\,2259+586 \citep{woods04} in
the 0.7--5 keV range. However, the high interstellar absorption
towards these objects causes a series of absorption edges in the
observed spectrum at low energies, introducing large systematic
uncertainties in the search for features in the intrinsic AXP
spectrum.
A hint of a spectral feature at $\sim$0.9 keV was noted by
\cite{durant06} in the spectrum of the AXP 4U\,0142+61, after its
deconvolution from interstellar absorption using the edges directly
observed in the high resolution X--ray spectra, but also this
measure is affected by the poorly constrained abundances of most
interstellar elements.
As shown in Figure \ref{lines}, for \src\ we did not find evidence
for lines, but, although this source is one of the dimmest AXPs, we
could put stringent limits on narrow features
in the soft X--ray band, which are virtually independent of the
photoelectric absorption model. The dipolar magnetic field derived
from the spin-down rate of \src\ is \mbox{$4\times10^{14}$ G},
corresponding to a proton cyclotron energy of $\sim$2.5 keV;
however, a cyclotron line at lower energies is expected if the
cyclotron emission or absorption process occurs far from the neutron
star surface, while a feature at higher energies is produced if
strong multipolar magnetic field components are present. These
effects, in addition to other processes that can suppress the
spectral features \citep[see, e.g.,][]{ho03}, make the lack of
proton cyclotron lines in the X--ray spectra of magnetars compatible
with their magnetic fields of (6--$250)\times10^{13}$ G
(corresponding to proton cyclotron energies of 0.4--15 keV) derived
from their timing properties.
\acknowledgements

We thank Nanda Rea for fitting the spectrum with the RCS model and
useful discussion. This research has been partially supported by the
Italian Space Agency.
\newpage
\bibliographystyle{apj}
\bibliography{biblio}

\end{document}